\documentclass[12pt]{article}
\usepackage[T1]{fontenc}
\usepackage{lmodern}
\usepackage[margin=2.5cm]{geometry}
\usepackage{amsmath,amssymb,amsthm}
\usepackage{booktabs}
\usepackage{adjustbox} 

\usepackage[authordate, backend=biber, natbib=true]{biblatex-chicago}
\usepackage{booktabs}

\usepackage{microtype}
\usepackage{xcolor}
\usepackage{hyperref}
\usepackage[normalem]{ulem} 
\usepackage{graphicx}
\definecolor{mycustomblue}{HTML}{023C85}

\makeatletter
\renewcommand{\fnum@figure}{\textbf{Figure \thefigure}}
\renewcommand{\fnum@table}{\textbf{Table \thetable}}
\makeatother

\usepackage{caption}
\hypersetup{
    colorlinks=true,
    linkcolor=mycustomblue,
    filecolor=mycustomblue,      
    urlcolor=mycustomblue,
    citecolor=mycustomblue,
}

\newcommand{\vct}[1]{\boldsymbol{#1}}
\newcommand{\mat}[1]{\boldsymbol{#1}}
\newcommand{\vtheta}{\boldsymbol{\theta}}
\newcommand{\vzeta}{\boldsymbol{\zeta}}
\newcommand{\vpsi}{\boldsymbol{\psi}}
\newcommand{\veta}{\boldsymbol{\eta}}
\newtheorem{theorem}{Theorem}[section]
\newtheorem{proposition}[theorem]{Proposition}
\newcommand{\R}{\mathbb R}
\newcommand{\Pois}{\operatorname{Poisson}}
\newcommand{\Normal}{\mathcal N}
\newcommand{\op}{o_{\mathsf P}}
\newcommand{\Op}{O_{\mathsf P}}

\newcommand{\dto}{\xrightarrow{\;\;d\;\;}}
\newcommand{\pto}{\xrightarrow{\;\,\mathsf{P}\;\,}}
\newcommand{\pos}[1]{\left[#1\right]_+}

\newcommand{\myemail}{silca\,[at]\,sam.sdu.dk}   

\title{
    Boundary inference for the Makeham term in mortality models}
\author{Silvio C. Patricio\thanks{Corresponding author. Email:
        \texttt{\myemail}. ORCID: \texttt{0000-0002-1025-5571}} \\[3pt]
        \small Interdisciplinary Center on Population Dynamics\\
        \small University of Southern Denmark \\
        }
\date{}

\begin{document}

\vspace{-3mm}

\maketitle

\vspace{-10mm}

\begin{abstract} \noindent
Whether a mortality model needs a Makeham term, the constant representing background mortality, is commonly decided with a likelihood-ratio test. The constant cannot be negative, so testing its absence places it on a boundary, where the usual chi-squared calibration fails. Assuming independent Poisson death counts and standard regularity conditions, we show that when the constant is the only parameter on a boundary the statistic converges to an equal mixture of a point mass at zero and a chi-squared distribution with one degree of freedom: at the $5\%$ level the critical value is $2.71$, not $3.84$. We state conditions under which the correction holds for Makeham models, verify them for Gompertz--Makeham and gamma-Gompertz--Makeham, and quantify the information about the constant, which fixes the local power, the smallest detectable term, and the loss of detectability as the age window narrows. When a gamma-frailty variance is estimated at a true value of zero, the limit is no longer a chi-squared mixture and the correction rejects too often. Monte Carlo experiments show the conventional cutoff rejecting at half the nominal level, a correctly calibrated test missing a term at its detection threshold in most samples, and excess rejections persisting as exposure grows.
\end{abstract}

\vspace{2mm}
\vspace{6mm}


\section{Introduction}

In Makeham mortality models the force of mortality splits into an age-dependent component and a non-negative constant, and fitting one requires deciding whether the data support that constant. The Gompertz slope describes the exponential increase of the age-dependent component and is widely read as a proxy for the rate of aging. A constant raises mortality by the same amount at every age, so its share of the force of mortality is largest where mortality is lowest. Omitting a constant that belongs leaves the age-dependent component to absorb background risk at the young end of the fitted range, which flattens the fitted slope, and the distortion grows as the range extends to younger ages \citep{patricio2026rhythm}.

\citet{makeham1860law} introduced the constant by adding it to the exponentially increasing force of mortality of the Gompertz law. It is usually interpreted as background or extrinsic risk, the part of the force of mortality often associated with accidents and infections rather than with senescence, although those causes need not themselves have age-invariant rates. \citet{patricio2024makeham} take the Makeham term as given and quantify its contribution, through the share of deaths it accounts for at each age and the threshold age after which senescent mortality predominates. The question here is the prior one, whether the data support a positive Makeham term at all.

Setting the constant to zero reduces Gompertz--Makeham to Gompertz, so the two models are nested and the decision is commonly made with a likelihood-ratio test. \citet{pletcher1999model} uses nested likelihood-ratio comparisons on age-specific mortality data to decide which components a model needs. Such comparisons support conclusions only if the reference distribution of the test statistic is correct.

We denote the Makeham parameter by $\kappa$ and test
\begin{equation*}
 H_0 : \kappa = 0, \qquad
 H_1 : \kappa>0.
 \end{equation*}
We restrict $\kappa\geq0$, so $H_0$ places it at the boundary of that range rather than in its interior. The usual result that the likelihood-ratio statistic has an asymptotic chi-squared distribution relies on the restricted parameter lying in the interior, and it fails here \citep{chernoff1954distribution,self1987asymptotic}.

The boundary problem is familiar in mortality research, and it has been worked out most fully for the frailty variance of the gamma-Gompertz model, where testing for mortality deceleration means testing whether that variance is zero \citep{bohnstedt2019detecting,bohnstedt2021information}. The same limit appears when a Gompertz hazard is tested against a constant hazard at extreme ages \citep{belzile2021human}. For the Makeham term, \citet{danko2018population} halve the likelihood-ratio $p$-values in a gamma-Gompertz--Makeham model because the null lies on the boundary of the parameter space. This correction requires the frailty variance to be an interior nuisance parameter. If its true value is zero and we estimate it under both hypotheses, testing the Makeham term involves two boundary constraints.

When $\kappa$ is the only parameter on a boundary and the regularity conditions in Appendix~\ref{app:regularity} hold, the test statistic converges to $\tfrac12\delta_0+\tfrac12\chi^2_1$ rather than to $\chi^2_1$, where $\delta_0$ denotes a point mass at zero. Half of that limiting probability lies at zero, so a test calibrated against $\chi^2_1$ at a nominal $5\%$ level has an asymptotic type~I error rate of $2.5\%$.

In practice the statistic should be compared with $2.71$, the $90$th percentile of $\chi^2_1$, rather than with $3.84$, or equivalently the $p$-value from $\chi^2_1$ should be halved when the statistic is positive. The conventional cutoff rejects less often under both hypotheses and misses some positive terms that the corrected test would detect. Every rejection it produces survives the correction; only some decisions to retain a Gompertz model can change.

The half-and-half limit for the Makeham term follows from the theory of \citet{self1987asymptotic} once its conditions hold, and it is the starting point here rather than its main result. We make three further contributions. First, we state the conditions under which the limit holds for Poisson death counts and Makeham mortality models, and verify the local information condition for the Gompertz and gamma-Gompertz hazards, keeping global identification as a separate assumption. Second, we quantify the information available for detecting a Makeham term. Its information is the variation left after the age-dependent component has absorbed what it can, and it determines the local power of the test, an approximation to the smallest detectable Makeham term, and how quickly detectability is lost as the age window narrows. Third, we show that when a gamma-frailty variance is estimated and its true value is zero, testing the Makeham term becomes a different nonregular problem. The limit is then not generally a mixture of chi-squared distributions, and the usual correction can make the test reject too often. Monte Carlo experiments illustrate the limits at finite exposures.

\section{Testing for a Makeham term}
\label{sec:test}

We observe death counts $D_i$ in age intervals $[x_i,x_i+1)$, for $i = 1,\ldots,m$, and treat the exposure $e_i>0$ in the same interval as fixed and expressed in person-years. We write the population force of mortality as
\begin{equation}
  \bar{\mu}(x;\kappa,\vtheta) = \kappa+h(x,\vtheta),\qquad\kappa\geq0,
 \label{eq:generalhazard}
\end{equation}
where $\kappa$ is the Makeham constant and $h(x,\vtheta)$ is an age-dependent component, which may be Gompertz, gamma-Gompertz, the age-dependent part of Siler, or another mortality law. We write $h_i(\vtheta) = h(x_i,\vtheta)$ for the force of mortality at the fitted ages and treat $ \bar{\mu}(x_i;\kappa,\vtheta)$ as the rate applying throughout $[x_i,x_i+1)$, the usual piecewise-constant approximation for single-year intervals. The vector $\vtheta$ represents the parameters of $h$, and where one of them is restricted to be positive, such as the Gompertz level $\lambda$, we take its logarithm, so that the true $\vtheta_0$ is interior to the fitted parameter set, as the conditions in Appendix~\ref{app:regularity} require.

Conditional on the exposures, we assume independent death counts with
\begin{equation*}
 D_i\sim\Pois\left(e_i\, \bar{\mu}(x_i;\kappa,\vtheta)\right),
\end{equation*}
which \citet{brillinger1986biometrics} and \cite{Macdonald_Richards_Currie_2018} motivate as an approximation for death counts given exposure. The likelihood is
\begin{equation*}
 L(\kappa,\vtheta) = \prod_{i = 1}^m
 \frac{\{e_i \bar{\mu}(x_i;\kappa,\vtheta)\}^{D_i}}{D_i!}\,
 \exp\left\{-e_i \bar{\mu}(x_i;\kappa,\vtheta)\right\},
\end{equation*}
and, dropping factors free of the parameters, the log-likelihood is
\begin{equation}
 \ell(\kappa,\vtheta) = \sum_{i = 1}^m
 \left[D_i\log\{\kappa+h_i(\vtheta)\}-e_i\{\kappa+h_i(\vtheta)\}\right].
 \label{eq:generalll}
\end{equation}
Maximum likelihood estimates follow from maximizing Eq.~\eqref{eq:generalll}, and both hypotheses are fitted to the same data.

\subsection{The statistic and its distribution under the null hypothesis}
\label{sec:null}

Let $\widehat{\vtheta}_0$ be the maximizer of Eq.~\eqref{eq:generalll} when $\kappa = 0$, and let $(\widehat\kappa,\widehat{\vtheta})$ be the maximizer over $\kappa\geq0$ and $\vtheta$. The statistic is
\begin{equation}
 T = 2\left\{\ell(\widehat{\kappa},\widehat{\vtheta})-\ell(0,\widehat{\vtheta}_0)\right\},
 \label{eq:generallrt}
\end{equation}
which measures the improvement the Makeham parameter provides in fit to the death counts. The restricted model is a special case of the full one, so $T$ is never negative, and larger values indicate stronger evidence for the Makeham term.

The test can only work if a Makeham term leaves a pattern in the data that the age-dependent component cannot produce on its own. We state this requirement formally in Appendix~\ref{app:regularity}, as the condition that the Fisher information matrix be positive definite.

To obtain the null distribution we reduce the maximization in Eq.~\eqref{eq:generallrt} to one over $\kappa$ alone. Holding $\kappa\geq0$ fixed and maximizing over the remaining parameters gives the profile log-likelihood
\begin{equation*}
 \ell_p(\kappa) = \max_{\vtheta}\ell(\kappa,\vtheta).
\end{equation*}
$T$ is twice the maximum of $\ell_p(\kappa)-\ell_p(0)$ over $\kappa\geq0$, which is why the quadratic approximation in Eq.~\eqref{eq:gmquadratic} is maximized over the same range. We define the profile score $U$ for $\kappa$ as the slope of the profile log-likelihood at $\kappa = 0$,
\begin{equation*}
 U = \ell_p'(0),
\end{equation*}
and the observed profile information $J$ as its negative curvature there,
\begin{equation*}
 J = -\ell_p''(0).
\end{equation*}
Both derivatives are taken from the right. A positive $U$ means that a small increase in $\kappa$ initially improves the fit, and both quantities already account for the estimation of $\vtheta$.

The profile score is a weighted sum, over ages, of the differences between observed and expected deaths, adjusted for having estimated $\vtheta$. It is approximately normal when exposures are large, and we make this precise in Appendix~\ref{app:proof}.

For the asymptotic argument we write $e_{i,n} = n\bar e_i$, where $\bar e_i>0$ are fixed baseline exposures and $n$ is a dimensionless multiplier common to all ages, so that total person-years grow while the exposure proportions across ages stay fixed. We call this setting \emph{proportional exposure growth}. For cohort data it corresponds to a larger birth cohort with the same survival pattern, and for period data to a larger population with the same age structure. We show in Appendix~\ref{app:profile} that $J$ is positive with probability tending to one and that
\begin{equation}
 Z = \frac{U}{\sqrt J} \dto\Normal(0,1).
 \label{eq:zlimit}
\end{equation}

Under $H_0$ the estimator $\widehat{\kappa}$ is of order $n^{-1/2}$. Setting $\kappa = u/\sqrt n$ with $u\geq0$ fixed, a second-order Taylor expansion of the profile log-likelihood around $\kappa = 0$ gives
\begin{equation}
 \ell_p(\kappa)-\ell_p(0)
  = \kappa U-\tfrac12\kappa^2J+\op(1),
 \label{eq:gmquadratic}
\end{equation}
where $\op(1)$ denotes a random remainder converging to zero in probability, uniformly over bounded $u$; \citet{van2000asymptotic} develops the underlying asymptotic theory.

Under $H_0$ the model has no Makeham term, but random variation in the death counts will push the profile score above zero about half the time. When $U\leq0$ the quadratic in Eq.~\eqref{eq:gmquadratic} is maximized at $\kappa = 0$; when $U>0$ it is maximized at $\kappa = U/J$. Doubling the maximum,
\begin{equation}
 T = \pos{Z}^{\,2}+\op(1),
 \label{eq:gmpositivepart}
\end{equation}
where $[z]_+ = \max(z,0)$. The maximized quadratic and profile gains differ by $\op(1)$, so Eq.~\eqref{eq:gmpositivepart} holds without the two maximizers coinciding (Appendix~\ref{app:boundary}). A standard normal variable is negative with probability one half, producing a zero statistic in the limit, and conditional on being positive its square is chi-squared with one degree of freedom. Half of the null distribution sits at a single point to which $\chi^2_1$ assigns no probability.

\begin{theorem}[Makeham-term test]
\label{thm:general}
Assume conditions (\ref{ass:parameter-space})--(\ref{ass:maximization}) in Appendix~\ref{app:regularity}, which require in particular that $\vtheta_0$ be interior to the fitted parameter set and that the Fisher information matrix be positive definite. Under $H_0$ and proportional exposure growth,
\begin{equation}
 T\dto\tfrac12\delta_0+\tfrac12\chi_1^2,
 \label{eq:generallimit}
\end{equation}
where $\delta_0$ denotes a point mass at zero.
\end{theorem}

We restrict the functional form of $h$ only through the conditions in Appendix~\ref{app:regularity}, so the correction is not specific to the Gompertz family and applies to any mortality law of the form in Eq.~\eqref{eq:generalhazard}. We verify the local information condition in Appendices~\ref{app:gm} and~\ref{app:ggm} for the two models of Sections~\ref{sec:gm} and~\ref{sec:ggm}; global identification remains a separate assumption. If a nuisance parameter itself lies at a bound, the theorem does not apply and a separate derivation is required; the case that arises in practice, a frailty variance whose true value is zero, is treated in Section~\ref{sec:zerofrailty}. The proof is in Appendix~\ref{app:proof}.

\subsection{The hypothesis-testing rule}
\label{sec:rule}

For a nominal level $\alpha\in(0,\tfrac12)$, we reject $H_0$ when
\begin{equation}
 T>\chi^2_{1,\,1-2\alpha},
 \label{eq:rule}
\end{equation}
where $\chi^2_{1,q}$ is the $q$ quantile of a chi-squared distribution with one degree of freedom. At $\alpha = 0.05$ the critical value is $2.7055$, compared with $3.8415$ under the usual degrees-of-freedom rule. For an observed value $t$ of the statistic, the asymptotic $p$-value is
\begin{equation}
 p(t) = 
 \begin{cases}
  \tfrac12 \mathbb P (V\geq t), & t>0,\\[2pt]
  1, & t = 0,
 \end{cases}
 \qquad V\sim\chi_1^2.
 \label{eq:pvalue}
\end{equation}
The value at $t = 0$ is the usual right-tailed $p$-value for a limiting distribution that places probability one half exactly at zero: a draw from that distribution is then at least as large as the observed statistic with probability one. For $t>0$ the rule amounts to halving the $p$-value obtained from $\chi^2_1$.

Rejection supports an age-independent background component in addition to the fitted age pattern. Non-rejection means the death counts do not provide sufficient evidence for that component over the fitted age range, and does not rule out an age-independent contribution to mortality. Both readings are conditional on the fitted age-dependent component, because the null hypothesis is that $\kappa=0$ within the chosen model and not that mortality has no age-independent part. We return to this in Section~\ref{sec:interpret}.

\subsection{Gompertz--Makeham model}
\label{sec:gm}

The Gompertz--Makeham model takes $h(x,\vtheta) = \lambda e^{\beta x}$, so that
\begin{equation}
  \bar{\mu}(x;\lambda,\beta,\kappa) = \lambda e^{\beta x}+\kappa,\qquad \lambda>0,\ \beta>0,\ \kappa\geq0.
 \label{eq:gmhazard}
\end{equation}
Under $H_0$ the constant $\kappa$ is zero and Eq.~\eqref{eq:gmhazard} reduces to the Gompertz model, while under $H_1$ a positive Makeham term is present. \citet{castellares2022gompertz} discuss the structural properties of this model. Substituting into Eq.~\eqref{eq:generalll} gives
\begin{equation*}
 \ell_{\mathrm{GM}}(\lambda,\beta,\kappa)
  = \sum_{i = 1}^m\left[D_i\log\{\lambda e^{\beta x_i}+\kappa\}-e_i\{\lambda e^{\beta x_i}+\kappa\}\right],
\end{equation*}
and the statistic Eq.~\eqref{eq:generallrt} compares the Gompertz--Makeham fit with the Gompertz fit obtained at $\kappa = 0$.

On the log scale the age-dependent component is the straight line $\log h(x) = \log\lambda+\beta x$, so $\lambda$ fixes its height and $\beta$ its slope. The question raised in Section~\ref{sec:null} becomes whether shifting or tilting that line can mimic adding a constant to the force of mortality itself, which lifts the log hazard most at younger ages and has diminishing relative influence as age increases. We show in Appendix~\ref{app:gm} that it cannot, once three distinct ages are observed with positive exposure.

\begin{proposition}[Gompertz--Makeham test]
\label{prop:gm}
Assume that the true $\lambda,\beta$ are strictly positive and that the identification and regularity conditions in Appendix~\ref{app:regularity} hold. Under $H_0$ and proportional exposure growth,
\begin{equation*}
 T\dto\tfrac12\delta_0+\tfrac12\chi_1^2.
\end{equation*}
\end{proposition}

\subsection{Gamma-Gompertz--Makeham model}
\label{sec:ggm}

Gamma frailty bends the age pattern at the opposite end of the age range from the Makeham term. As individuals with higher frailty tend to die earlier within a cohort, the population hazard rises more slowly with age than the individual hazard, which produces deceleration and eventually a plateau \citep{vaupel1979impact, missov2015plateau}. The Makeham constant instead lifts the log hazard at young adult ages. A test for $\kappa$ with $\gamma$ estimated asks whether the data need an age-independent component on top of what selection already accounts for at the older ages. This selection argument follows a cohort as it ages. A period schedule combines different cohorts, so a gamma-Gompertz fit to period rates describes their age pattern without, by itself, identifying frailty variation within a cohort.

We denote the gamma-frailty variance by $\gamma\geq0$ and define
\begin{equation}
 g(x;\lambda,\beta,\gamma) = \frac{\lambda e^{\beta x}}{1+\gamma\frac \lambda\beta(e^{\beta x}-1)}.
 \label{eq:gg}
\end{equation}
Taking $h(x,\vtheta) = g(x;\lambda,\beta,\gamma)$ in Eq.~\eqref{eq:generalhazard} gives the gamma-Gompertz--Makeham hazard
\begin{equation*}
  \bar{\mu}(x;\lambda,\beta,\gamma,\kappa) = \kappa+g(x;\lambda,\beta,\gamma),
\end{equation*}
in which frailty acts on the Gompertz component while the Makeham term remains additive. For $\gamma>0$, $g$ tends to $\beta/\gamma$ as age increases, so the constant here raises the plateau from $\beta/\gamma$ to $\kappa+\beta/\gamma$ instead of becoming negligible at the oldest ages, which is how this case differs from the Gompertz one. 

We test $H_0:\kappa = 0$ against $H_1:\kappa>0$, estimating $\gamma$ as a nuisance parameter; a positive true variance is treated here, and a zero one in Section~\ref{sec:zerofrailty}. If the true frailty variance $\gamma_0$ is strictly positive, $\kappa$ remains the only parameter on a boundary, because $\gamma$ can be perturbed in both directions and so behaves like an ordinary nuisance parameter. Theorem~\ref{thm:general} then applies with $\vtheta = (\log\lambda,\beta,\gamma)^\top$.

\begin{proposition}[Gamma-Gompertz--Makeham test with positive frailty variance]
\label{prop:ggm}
Under $H_0$, assume $\gamma_0>0$ and that the identification and Fisher information conditions in Appendix~\ref{app:regularity} hold for the nuisance vector $(\log\lambda,\beta,\gamma)^\top$. Under proportional exposure growth,
\begin{equation*}
 T\dto\tfrac12\delta_0+\tfrac12\chi_1^2.
\end{equation*}
\end{proposition}

Here the information condition asks whether changes in $\lambda,\beta,\gamma$ can reproduce a constant addition at every age. We give the derivatives in Appendix~\ref{app:ggm} and show that four distinct ages with positive exposures are enough, provided $\lambda\gamma\neq\beta$.

\subsection{When the frailty variance is also on the boundary}
\label{sec:zerofrailty}

A frailty variance of zero is a realistic possibility whenever the data show little deceleration, and whether frailty is needed is itself the subject of dedicated tests \citep{bohnstedt2019detecting,patricio2023using}. If $\gamma$ is fixed at zero in both fits, the test reduces to the Gompertz--Makeham case of Section~\ref{sec:gm}. If instead $\gamma\geq0$ is estimated and its true value is zero, both $\kappa$ and $\gamma$ lie on their boundaries under $H_0$, and the distribution of the statistic in Eq.~\eqref{eq:generallrt} must account for both constraints.

Tests of a variance component often run into the issue that the remaining parameters become unidentified when the variance is zero, which invalidates the usual expansions. That does not happen here. At $\gamma = 0$, Eq.~\eqref{eq:gg} reduces to the Gompertz hazard, whose parameters $\lambda$ and $\beta$ remain identified, and a small positive $\gamma$ changes the age pattern smoothly, so the usual expansions still apply. The problem is a boundary problem of the \citet{self1987asymptotic} type rather than a non-identifiability problem.

We derive the limiting statistic $T_0$ for the case $\kappa = \gamma = 0$ in Appendix~\ref{app:zero-frailty}. Its distribution depends on $\mat{K}$, the efficient Fisher information matrix for $(\kappa,\gamma)^\top$. Writing $F_{\mat{K}}$ for the distribution function of $T_0$, the rejection rule at a true frailty variance of zero is
\begin{equation}
 T>F_{\mat{K}}^{-1}(1-\alpha).
 \label{eq:zero-frailty-rule}
\end{equation}

The distribution of $T_0$ is not in general a mixture of chi-squared distributions. Under $H_0$ we still estimate $\gamma\geq0$, so the null parameter set is an edge of the quadrant rather than its origin, and that is what breaks the mixture form. When both parameters are fixed at zero under the null hypothesis, the limit is still such a mixture, with weights that depend on the information matrix \citep{self1987asymptotic,silvapulle2005constrained}. In particular, the critical value $2.7055$ from Section~\ref{sec:rule} does not carry over. For a general $\mat K$, $F_{\mat K}$ should be evaluated by simulation from the explicit expression for $T_0$ derived in Appendix~\ref{app:zero-explicit}, and we treat in Appendix~\ref{app:zero-negcorr} the special case in which the earlier mixture is recovered.

For the Gompertz hazard, changes in the Makeham term and frailty can partly compensate for one another after we fit the level and slope. A small $\kappa$ bends the log hazard upward at young ages, making it convex, whereas a small $\gamma$ bends it downward at old ages, making it concave. These opposite curvatures suggest a negative covariance $k_{12}$ between the two adjusted scores, which we confirm in Section~\ref{subsec:sim_frailty} for its simulation design. When $k_{12}<0$, we show in Appendix~\ref{app:zero-negcorr} that the limiting law places probability $\tfrac12+\arcsin(\rho)/(2\pi)<\tfrac12$ at zero, where $\rho=k_{12}/\sqrt{k_{11}k_{22}}$ is the correlation of the two adjusted scores and the arcsine term comes from Sheppard's classical formula for the orthant probabilities of a bivariate normal distribution \citep{sheppard1899application}, and that its upper tail lies above that of $\tfrac12\delta_0+\tfrac12\chi^2_1$ at every point. The rule in Eq.~\eqref{eq:rule} then rejects a true null hypothesis more often than its nominal level, at every level $\alpha$.

These results distinguish a true frailty variance of zero from a fixed positive one, but a fitted value cannot. A small $\widehat{\gamma}$ is compatible with a positive true variance, where Theorem~\ref{thm:general} applies, and with a zero one, where it does not, so choosing the critical value from $\widehat{\gamma}$ alone has no established guarantee of keeping the false-detection probability at $\alpha$. We show in Appendix~\ref{app:zero-near} why variances approaching zero require further analysis.

\subsection{Where the evidence about \texorpdfstring{$\kappa$}{kappa} comes from}
\label{sec:power}

A Makeham term raises mortality by the same absolute amount at every age. On the log scale this addition is large relative to the age-dependent hazard at young adult ages and shrinks as that hazard grows. Its share of the force of mortality at age $x$, $\kappa/\{\kappa+h(x,\vtheta)\}$, is the expected proportion of deaths at that age that the model attributes to the age-independent component \citep{patricio2024makeham}. Under a Gompertz hazard the share declines to zero at the oldest ages. Under gamma frailty it tends instead to $\kappa/(\kappa+\beta/\gamma)$, because the constant raises the plateau (Section~\ref{sec:ggm}). The information available about $\kappa$ depends also on the exposures and on the other parameters of $h$.

\subsubsection{Information after estimating the age-dependent component}

We write the efficient information for $\kappa$ in Appendix~\ref{app:regularity} as a weighted least-squares quantity in which age $i$ carries the weight $\bar e_i/h_i$. What the weights multiply, however, is the residual left after projecting the constant vector on the nuisance derivative directions, and that residual depends on $h$.

For the Gompertz component of Section~\ref{sec:gm}, $\vct r_i = (h_i,x_ih_i)^\top$, so the residual is $1-h_i(q_1+q_2x_i)$ at the minimizing coefficients $q_1,q_2$. These coefficients depend on the mortality schedule and exposures across all fitted ages, so a small $h_i$ does not by itself make the residual close to one, and the weights determine how much each age contributes.

For a fixed age range and exposure pattern, total expected efficient information at the null is $n\,j_{\mathrm{eff}}$. Whenever $j_{\mathrm{eff}}>0$, increasing exposure improves precision and the ability to detect small Makeham terms. Comparing age windows requires evaluating both the weights and the residuals in Eq.~\eqref{eq:projection}; the formula alone does not rank young and old ages by information per person-year.

\subsubsection{Local power and the detection threshold}

The same quantity determines the size of Makeham term the test can detect. \citet{bohnstedt2019detecting} derive the local power of the boundary likelihood-ratio test for the frailty variance of the gamma-Gompertz model, and the same argument applies to $\kappa$ (Appendix~\ref{app:power}). For a term of order $n^{-1/2}$, $\kappa=u/\sqrt n$, the rule in Eq.~\eqref{eq:rule} rejects with probability tending to $\Phi(u\sqrt{j_{\mathrm{eff}}}-z_{1-\alpha})$, where $\Phi$ is the standard normal distribution function and $z_{1-\alpha}$ its $1-\alpha$ quantile, so that $z_{0.95}^2=2.7055$. Setting this probability equal to a target power $\pi$ and solving for $\kappa$ gives a local approximation to the smallest Makeham term detected with that power,
\begin{equation*}
 \kappa_{\min}\approx\frac{z_{1-\alpha}+z_{\pi}}{\sqrt{n\,j_{\mathrm{eff}}}},
\end{equation*}
where $z_{\pi}$ is the $\pi$ quantile of the standard normal distribution. With $\alpha=0.05$ and the conventional target $\pi=0.8$ \citep{cohen1988statistical}, $z_{0.80}=0.84$ and $\kappa_{\min}\approx2.486/\sqrt{n\,j_{\mathrm{eff}}}$. The conventional critical value $3.8415$ raises the threshold and so lowers power \citep{bohnstedt2021information}. It replaces $z_{0.95}$ by $z_{0.975}=1.96$, and for a term of size $\kappa_{\min}$ the local-power approximation gives a rejection probability of about $0.70$ rather than $0.80$.

\subsubsection{The age window}

Power also depends on the age window. \citet{bohnstedt2021information} use the Fisher information to show how the power to detect mortality deceleration changes with the age range of the data, and the same reasoning applies to $\kappa$ through $j_{\mathrm{eff}}$. For the Gompertz hazard, narrowing the window has a strong effect. Over a short range of ages, a small change in the Gompertz level and slope reproduces a constant addition almost exactly, because near any age $x_0$ the combination $\{1-\beta(x-x_0)\}e^{\beta(x-x_0)}$ of $h$ and $xh$ equals $1-O\{(x-x_0)^2\}$ (Appendix~\ref{app:shortwindow}). Little is then left for $\kappa$ to explain, $j_{\mathrm{eff}}$ is small, and a non-rejection over a short window can reflect low power rather than a negligible Makeham term.
\subsection{What the statistic responds to}
\label{sec:interpret}

The null hypothesis of Theorem~\ref{thm:general} is that $\kappa=0$ within the chosen model, and the statistic compares two fits of that model. It responds to any systematic difference between the data and $h$ that a constant can partly absorb, not only to an age-independent component of mortality. A rejection or a non-rejection is conditional on $h$ describing the age pattern over the fitted window.

Departures at the young end of the window can move the statistic in either direction. Childhood mortality or an accident hump raises mortality above a Gompertz line at some young ages and not at others, so its effect on $\widehat\kappa$ depends on the window and on the shape of the excess. Fitting Gompertz--Makeham from age 20, \citet{patricio2024makeham} found the Makeham term underestimated rather than inflated in populations with an irregular age pattern between ages 20 and 45.

Old-age deceleration can conceal a Makeham contribution when we fit Gompertz--Makeham. The log hazard of Gompertz--Makeham is convex in age, because $\kappa$ lifts it most where the Gompertz component is small, whereas deceleration makes the observed log hazard concave at old ages. A positive $\kappa$ cannot reproduce that concavity. When deceleration dominates the fitted window, the constrained fit can place $\widehat\kappa$ at zero and give $T=0$, even when an age-independent component is present. The outcome depends on the age window, exposures, and mortality schedule. When the fitted window reaches ages with deceleration, the Makeham term should be tested within a model that allows for it, such as gamma-Gompertz--Makeham.

\section{Numerical illustrations}
\label{sec:simulations}

We use Monte Carlo experiments to illustrate the consequences of the reference distribution and available exposure for detecting a Makeham term. The simulations address three points: false detections when the Makeham term is the only parameter on a boundary, the power to detect a positive term, and the change in calibration when we also estimate a frailty variance whose true value is zero. The aim is illustration rather than validation of the asymptotic results established in Section~\ref{sec:test}.

\subsection{Finite-sample calibration}
\label{subsec:sim_calibration}

We first simulate death counts under a Gompertz hazard with no Makeham term,
\begin{equation*}
D_{i,n}\sim\operatorname{Poisson}
\left\{
n\bar e_i\lambda_0\exp(\beta_0x_i)
\right\},
\end{equation*}
for single years of age $x_i\in[30,90]$, with $\lambda_0=2\times10^{-5}$ and $\beta_0=0.10$. We take the baseline exposures proportional to the survivorship implied by the same schedule, $\bar e_i\propto\exp\{-(\lambda_0/\beta_0)(e^{\beta_0x_i}-e^{30\beta_0})\}$, and normalize them to sum to one, so that $n$ is the total exposure in person-years and the design introduces no parameter beyond the schedule itself. We vary $n\in\{10^4, 10^5, 10^6, 10^7\}$ while keeping these exposure proportions fixed, as in the asymptotic framework of Section~\ref{sec:test}. Throughout the numerical illustrations, an increase in exposure means an increase in this total $n$, not a change in its distribution across ages. For each value of $n$ we generate $B=10{,}000$ independent samples and fit the Gompertz and Gompertz--Makeham models.

\begin{table}[ht]
\centering
\caption{Finite-sample calibration of the Makeham likelihood-ratio test.}
\label{tab:sim_calibration}

\begin{tabular}{rrrrr}
\hline
$n$ & $\Pr(T \approx 0)$ & $\Pr(T > 2.7055)$
    & $\Pr(T > 3.8415)$ & Failed \\
\hline
$10^4$ & 0.566 & 0.041 & 0.021 & 0 \\
$10^5$ & 0.518 & 0.043 & 0.024 & 1 \\
$10^6$ & 0.511 & 0.052 & 0.027 & 3 \\
$10^7$ & 0.496 & 0.049 & 0.023 & 8 \\
\hline
Asymptotic & 0.500 & 0.050 & 0.025 & -- \\
\hline
\end{tabular}%

\par\smallskip
\parbox{.75\columnwidth}{%
  \scriptsize
  $B=10{,}000$ per exposure level. Probabilities use successful
  fits; failures are shown separately. We classify
  $|T_{\mathrm{raw}}| \leq 1 \times 10^{-7}$ as numerical zero.
}
\end{table}

We compare in Table~\ref{tab:sim_calibration} the empirical rejection probabilities under the boundary-corrected cutoff $2.7055$ and the conventional $\chi^2_1$ cutoff $3.8415$. The corrected test rejects slightly less often than the nominal $5\%$ level at the two smaller exposures, with rejection probabilities of $0.041$ and $0.043$. At the two larger exposures, these probabilities are $0.052$ and $0.049$, close to the nominal level. Using the conventional cutoff instead gives rejection probabilities between $0.021$ and $0.027$, consistent with its asymptotic rejection probability of $0.025$.

A small number of fits failed to converge. The count rises with exposure, which is consistent with a sharper likelihood and fixed convergence tolerances, and at no more than eight failures in $10{,}000$ samples it does not affect the reported probabilities. We classify a statistic as numerical zero when $|T_{\mathrm{raw}}|\leq10^{-7}$, where $T_{\mathrm{raw}}$ is twice the fitted log-likelihood difference before numerical truncation at zero. The proportion of such statistics approaches one half as $n$ increases, consistent with the limiting law
\begin{equation*}
\frac12\delta_0+\frac12\chi^2_1.
\end{equation*}
For this design, the boundary correction provides approximately nominal calibration at the larger exposures, whereas the conventional cutoff remains conservative.

Increasing exposure improves the accuracy of the boundary approximation in this design, but does not make the conventional cutoff appropriate for a $5\%$ test. At the smaller exposures the corrected test still rejects somewhat less often than its nominal level, and we return in the Discussion to whether a higher-order correction could close that gap. Rejecting at the intended rate under the null does not, however, tell us whether a positive component is large enough to be detected.
\subsection{Local power and detectability}
\label{subsec:sim_power}

We next measure power: the probability that the test detects a Makeham term when a positive term is present. We generate counts under local alternatives
\begin{equation*}
\kappa_n=\frac{u}{\sqrt n},
\end{equation*}
with $u>0$, so the positive constant approaches zero as total exposure $n$ grows. We evaluate the efficient information $j_{\mathrm{eff}}$ under the null using Eq.~\eqref{eq:projection} and express the alternatives through the standardized shift
\begin{equation*}
\delta=u\sqrt{j_{\mathrm{eff}}},
\end{equation*}
which combines the size of the Makeham term with the information available to detect it after estimating the age-dependent component. The local-power approximation is
\begin{equation*}
\pi(\delta)=\Phi\left(\delta-z_{1-\alpha}\right).
\end{equation*}
At a fixed $\delta$, the Makeham constant decreases as $n$ increases. We assess the detection of progressively smaller age-independent contributions to mortality, rather than hold the mortality schedule fixed across exposure levels.

\begin{figure}[ht!]
    \centering
    \includegraphics[width=.9\columnwidth]{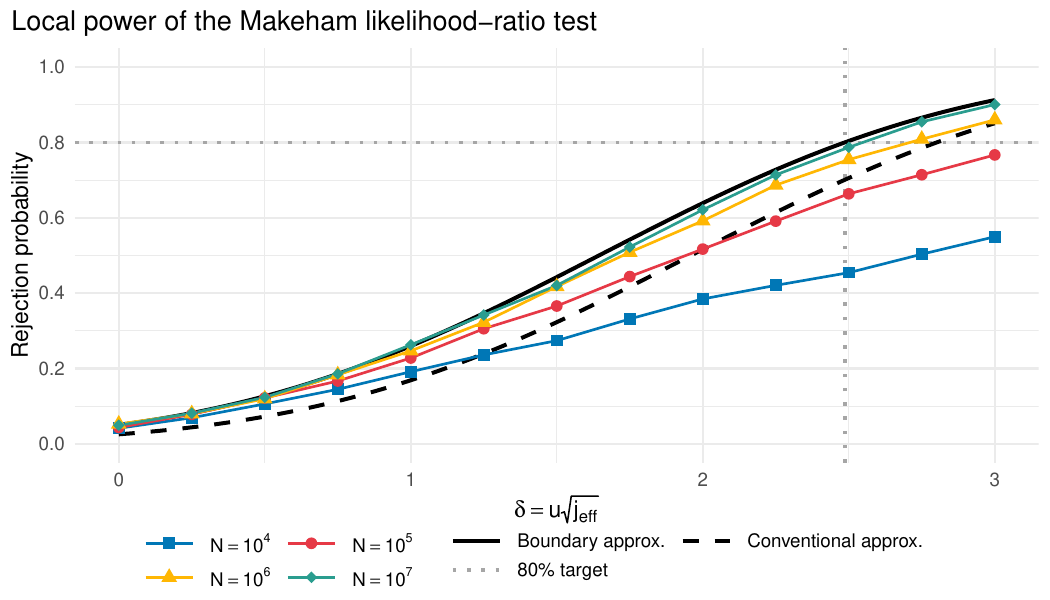} \caption{Local power of the Makeham likelihood-ratio test. Colored curves show empirical rejection probabilities using the boundary-corrected cutoff $2.7055$ at each exposure level. The solid black curve shows the local-power approximation for the corrected cutoff; the dashed black curve shows it for the conventional cutoff. The dotted lines mark the target power of $0.80$ and its corresponding standardized shift, $\delta\approx2.486$.}
    \label{fig:sim_power}
\end{figure}

Figure~\ref{fig:sim_power} shows that empirical power approaches the local approximation as $n$ increases. For this design, the approximation overstates power, particularly at the smallest exposure. Part of that gap is a difference in level rather than in power, because at $n=10^4$ the corrected test rejects in $4.1\%$ of null samples rather than $5\%$ (Table~\ref{tab:sim_calibration}), so the same samples are judged against an effectively stricter test than the approximation assumes.

At $\alpha=0.05$ and target power $\pi=0.80$, the approximate detection threshold of Section~\ref{sec:power} is
\begin{equation*}
\kappa_{\min}
\approx
\frac{z_{0.95}+z_{0.80}}{\sqrt{n j_{\mathrm{eff}}}}.
\end{equation*}
For this design, $j_{\mathrm{eff}}\approx243.96$. At the corresponding thresholds, empirical power using the corrected cutoff increases from approximately $46\%$ at $n=10^4$ to $78\%$ at $n=10^7$, approaching the target of $80\%$. Applying the conventional cutoff to the same simulated samples lowers power by approximately $11$ percentage points across all four exposure totals.

These results show why failure to detect a Makeham term need not imply that its contribution to mortality is negligible. At the smallest exposure, the corrected test fails to detect the positive term in more than half the simulated samples, even though the local approximation predicts $80\%$ power. A decision to retain the Gompertz model can reflect limited information rather than the absence of an age-independent component. 

The boundary correction reduces missed detections but does not remove this dependence on the information available, so a non-rejection is best read alongside the size of the term the data could have detected, taking $\kappa_{\min}$ as an approximate guide whose accuracy improves with exposure.

When comparing mortality schedules across populations or periods with different exposures, we must also account for differences in detectability. A Makeham term may be supported in one dataset and remain undetected in another because the available information differs. Such differences in test outcomes do not, by themselves, establish differences in the underlying mortality components.

\subsection{The joint boundary at zero frailty}
\label{subsec:sim_frailty}

Finally, we consider the setting of Section~\ref{sec:zerofrailty}, in which both the Makeham constant and the true frailty variance are zero. We generate counts from the same Gompertz schedule with
\begin{equation*}
\kappa_0=\gamma_0=0.
\end{equation*}
This generating point belongs to the null hypothesis $\kappa=0$, which leaves the frailty variance unknown. Under the null we fit gamma-Gompertz with $\kappa=0$ and estimate $\gamma\geq0$. Under the alternative, we fit gamma-Gompertz--Makeham and estimate both $\kappa\geq0$ and $\gamma\geq0$. We test for the Makeham term alone, while allowing for frailty under both hypotheses. Fixing $\gamma=0$ in both fits would instead recover the Gompertz--Makeham test of Section~\ref{subsec:sim_calibration}.

For this design, the efficient information matrix $\mathbf K$ gives the adjusted-score correlation
\begin{equation*}
\rho
=
\frac{k_{12}}{\sqrt{k_{11}k_{22}}}
\approx -0.630.
\end{equation*}
The negative correlation reflects the interaction between the two components after accounting for the Gompertz level and slope. As we show in Appendix~\ref{app:zero-frailty}, this interaction changes the null distribution of the Makeham test.

\begin{figure}[ht!]
    \centering
    \includegraphics[width=.9\linewidth]{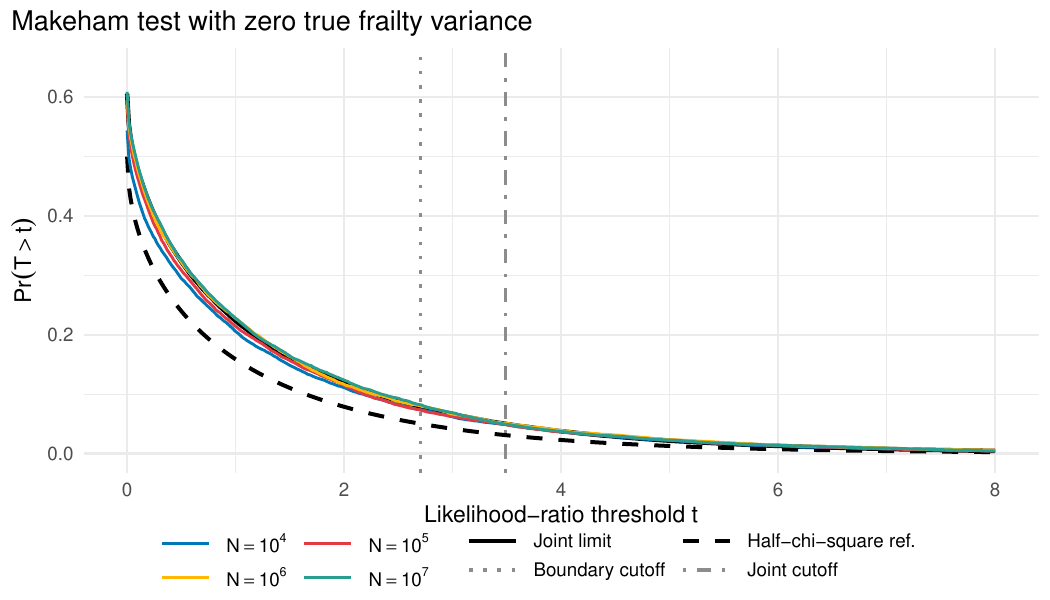}
    \caption{Null distribution of the Makeham likelihood-ratio statistic when the true frailty variance is zero. Colored curves show finite-sample upper-tail probabilities at each exposure level. The solid black curve shows the simulated joint-boundary limit, and the dashed black curve shows the half-and-half chi-squared reference distribution. The vertical lines mark the usual boundary cutoff $2.7055$ and the simulated joint-boundary cutoff $3.489$. Frailty remains estimated under both hypotheses.}
    \label{fig:sim_frailty}
\end{figure}

We compare the finite-sample statistic with the joint-boundary limit by drawing $100{,}000$ realizations of
\begin{equation*}
\mathbf W\sim\mathcal N_2(\mathbf 0,\mathbf K)
\end{equation*}
and evaluating $T_0(\mathbf W)$. Figure~\ref{fig:sim_frailty} shows the corresponding upper-tail probabilities. At any threshold $t$, the height of a curve gives the probability of rejecting the null when we use $t$ as the critical value.

The finite-sample curves approach the joint-boundary limit as exposure increases. The proportion of statistics classified as numerical zero decreases from $0.456$ at $n=10^4$ to $0.391$ at $n=10^7$, close to the theoretical value
\begin{equation*}
\Pr(T_0=0)
=
\frac12+\frac{\arcsin(\rho)}{2\pi}
\approx0.391.
\end{equation*}
The usual half-and-half law instead assigns probability $0.5$ to zero and understates the upper-tail probabilities for this design.

This difference affects the frequency with which we infer a Makeham component. The cutoff $2.7055$ leaves $5\%$ of the half-and-half reference distribution above it, but approximately $7.9\%$ of the joint-boundary limit. With this cutoff, the empirical rejection probability should approach $0.079$, not $0.05$, as exposure increases. The simulated rejection probabilities, between $0.073$ and $0.082$ across the four exposure totals, follow this pattern. The test detects a Makeham term in roughly eight out of every hundred samples, although the generating schedule contains none and the nominal level allows five. Increasing exposure cannot remove this excess because it arises from the wrong limiting distribution.

The simulated $95$th percentile of the joint-boundary law is $3.489$. Using this cutoff gives empirical rejection probabilities between $0.049$ and $0.051$, close to the nominal level. Here we calculate $\mathbf K$ from the known generating parameters, so this comparison isolates the effect of the reference distribution. It does not establish the performance of a procedure that estimates $\mathbf K$ or selects a calibration from the fitted frailty variance.

Allowing for frailty changes more than the shape of the fitted mortality schedule: it can also change the evidence required to support an additional age-independent component. When the true frailty variance is zero but we estimate it, the usual Makeham boundary correction can mistake random variation for evidence of that component too often. These results distinguish this setting from one in which we fix frailty at zero before fitting, and they do not justify choosing between reference distributions solely according to whether the fitted variance equals zero.

\section{Conclusion}

When the Makeham constant is the only parameter on a boundary, the likelihood-ratio statistic for its absence converges to $\tfrac12\delta_0+\tfrac12\chi_1^2$. The practical rule follows directly. At the $5\%$ level the statistic should be compared with $2.71$, or the $p$-value from $\chi^2_1$ halved when the statistic is positive, while using the $95$th percentile of $\chi^2_1$, $3.84$, lowers both the asymptotic type I error rate and the power of the test. The result holds for Makeham models of the form $\kappa+h(x,\vtheta)$ under the stated conditions, and we have verified its local information condition for Gompertz--Makeham and gamma-Gompertz--Makeham.

The calibration is only part of what determines a conclusion. The information about $\kappa$ is the variation left after the age-dependent component has absorbed what it can, so the same test can be decisive on a wide age window and nearly powerless on a narrow one, and a non-rejection has to be read against the approximate size of the Makeham term the test could detect at a specified power. A correctly calibrated test can still be badly underpowered. At the smallest exposure considered here it misses a term at its own nominal detection threshold in more than half of the samples, so retaining a Gompertz model may record the limits of the data rather than the absence of a background component.

Two further conditions govern what a rejection means. The test is conditional on the fitted model, because old-age deceleration fitted with Gompertz--Makeham can push the estimate of $\kappa$ to zero. And when a gamma-frailty variance is estimated at a true value of zero, the half-and-half correction is not generally valid and rejects too often whenever the two adjusted scores are negatively correlated. That excess does not diminish as exposure grows, because it comes from using the wrong limiting distribution rather than from a finite-sample approximation.

Three assumptions bound these results. They are asymptotic, under proportional growth of exposures across ages. The two frailty cases treated here also leave out a variance that is unknown and arbitrarily close to zero, for which keeping the rejection probability at $\alpha$ requires an argument holding uniformly across such sequences.

The shortfall at the smaller exposures in Table~\ref{tab:sim_calibration} is of the kind that higher-order corrections are designed to remove. \citet{monti2018adjustments} show that the likelihood-ratio statistic remains Bartlett adjustable when the tested parameter lies on the boundary, so that an adjusted statistic approximates the mixture limit to order $o(n^{-1})$ rather than $O(n^{-1})$. Obtaining the adjustment factor here would require the third- and fourth-order cumulants of the profile score under the Poisson model, and their condition that the third cumulant of the score vanish would have to be verified rather than assumed. Their results also cover a single constrained parameter, so they do not extend to the joint boundary of Section~\ref{sec:zerofrailty}. We leave the adjustment for future work, noting that the exposures at which it would matter are small relative to those of national mortality data.

We have also assumed independent Poisson death counts, the standard sampling model in mortality estimation, although real counts are sometimes overdispersed relative to Poisson \citep{dean2016overdispersion, Hilbe2007NegativeBR}. The half-and-half limit comes from the constraint $\kappa\geq0$ rather than from the Poisson assumption, so the boundary argument itself would carry over to a count model that accommodates overdispersion, such as the Bell distribution \citep{castellares2018bell}, which \citet{castellares2024gammagompertzmakeham} use to fit the gamma-Gompertz--Makeham model. The remaining steps of the proof would have to be established again for that model, because consistency, the score limit, and the information identity all rest on the Poisson likelihood.

Deciding whether a mortality schedule needs a Makeham term takes three things at once: the correct reference distribution, an age window wide enough to carry information about the constant, and a model for the age-dependent component that matches the data over that window. The test supplies none of them on its own. It returns a number whose meaning depends on all three, and the first of them, the one that is purely a matter of calculation, is the one most often got wrong. What follows from the decision reaches beyond the constant itself. When the term is dropped, the Gompertz component takes up what it would have carried, and the fitted rate of aging moves with it. Calibrating this test is not a refinement at the margin of a mortality fit; it bears on the quantity that mortality models are most often fitted to estimate.

\section*{Acknowledgments}
The author thanks Fredy Castellares and Paola Vazquez-Castillo for their helpful comments and feedback on an earlier version of this work.

\paragraph*{Funding:}
This work was funded by the European Union (ERC, Born Once -- Die Once, Grant agreement ID 101043983). Views and opinions expressed are those of the author only and do not necessarily reflect those of the European Union or the European Research Council Executive Agency. Neither the European Union nor the granting authority can be held responsible for them.

\paragraph*{Author Contributions:}
S.C.P. is the sole author and contributed to all aspects of the work, including Conceptualization, Methodology, Software, Validation, Formal analysis, Investigation, Writing -- original draft, Writing -- review and editing, and Visualization.

\paragraph*{Disclosure statement:} 
The author reports there are no competing interests to declare.

\paragraph*{Data Availability Statement:}
The code, simulation settings, and materials required to reproduce the analyses, tables, and figures are openly available at \url{https://github.com/scpatricio/Makeham_LRT}.



\numberwithin{equation}{section}

\appendix

\section*{\Large Proofs and technical results}
The appendices collect the material supporting Section~\ref{sec:test}: the asymptotic framework and regularity conditions, the proof of Theorem~\ref{thm:general} and the local power derived from it, the local information conditions for the two models treated there, and the case of a frailty variance estimated at a true value of zero, where two parameters sit on boundaries at once.

\section{Assumptions and Fisher information}
\label{app:regularity}

We keep the $m$ age intervals fixed and increase their exposures proportionally. For fixed baseline exposures $\bar e_i>0$, we define
\begin{equation}
 e_{i,n} = n\bar e_i. 
 \label{eq:exposure}
\end{equation}
This is a mathematical convenience for the asymptotics. In practice the exposures are fixed, and $n$ represents the scale of the population. We write $\vtheta\in\R^p$ for the vector of $p$ parameters of the age-dependent component and $h_i(\vtheta)=h(x_i,\vtheta)$. Conditional on these exposures, we assume independent random death counts with
\begin{equation*}
 D_{i,n}\sim\Pois\left(n\bar e_i\{\kappa+h_i(\vtheta)\}\right).
\end{equation*}
Every likelihood in the note follows by specifying $h_i$ in this model. We write $ \bar{\mu}_i(\kappa,\vtheta) = \kappa+h_i(\vtheta)$ for the population force of mortality at age $x_i$ and take $n\to\infty$. Total exposure equals $n\sum_i\bar e_i$, so $n$ scales total person-years without changing their age distribution.

We use the following sufficient conditions for Theorem~\ref{thm:general}:
\begin{enumerate}
\renewcommand{\theenumi}{A\arabic{enumi}}
\renewcommand{\labelenumi}{(\theenumi)}
\item\label{ass:parameter-space} We fit the model over $[0,\bar\kappa]\times\mathcal T$, where $\bar\kappa>0$ and $\mathcal T\subset\R^p$ is compact. The set $\mathcal T$ contains an open ball around the true $\vtheta_0$. $\vtheta$ can be varied in every direction by a sufficiently small amount, while the constraint $\kappa\geq0$ remains binding at $\kappa = 0$.
\item\label{ass:identification} At every fitted age, $\bar\mu_i(\kappa,\vtheta)$ depends continuously on the parameters, so small parameter changes produce small changes in the force of mortality. Two distinct parameter vectors cannot give the same values at all fitted ages. There is also a constant $\varepsilon>0$ such that $\bar\mu_i(\kappa,\vtheta)\geq\varepsilon$ for every fitted age and every parameter vector in the fitted set.
\item\label{ass:smoothness-information} Each $h_i(\vtheta)$ has continuous partial derivatives with respect to $\vtheta$ up to order three near $\vtheta_0$. The normalized Fisher information matrix in Eq.~\eqref{eq:information} is positive definite.
\item\label{ass:maximization} Under each hypothesis, we use a parameter vector that gives the largest log-likelihood over its entire fitted parameter set. A numerical approximation is sufficient if the difference between that largest value and the value returned by the optimizer converges to zero in probability, that is, is $\op(1)$.
\end{enumerate}
Compactness is used to control the maximizers; suitable consistency conditions can replace it on an unbounded parameter set.

Under $H_0$, we write $h_i = h_i(\vtheta_0)$ and define the fixed derivative vector
\begin{equation*}
 \vct r_i = \left.\frac{\partial h_i(\vtheta)}{\partial\vtheta}\right|_{\vtheta = \vtheta_0}.
\end{equation*}
Differentiating Eq.~\eqref{eq:generalll} gives the scalar score for $\kappa$ and the vector score for $\vtheta$,
\begin{equation*}
 U_\kappa = \sum_i\left(\frac{D_{i,n}}{h_i}-n\bar e_i\right), \qquad
 \vct U_\theta = \sum_i\left(\frac{D_{i,n}}{h_i}-n\bar e_i\right)\vct r_i,
\end{equation*}
both of which are weighted sums of the centered residuals $D_{i,n}-n\bar e_ih_i$, which yields the normality established in Appendix~\ref{app:proof}.

We define Fisher information as the expected negative Hessian of the log-likelihood; it measures the expected curvature around the true parameter \citep[Chapter~7]{van2000asymptotic}. Dividing by $n$ gives the fixed matrix
\begin{equation}
 \mat I = \sum_i\frac{\bar e_i}{h_i}
 \begin{pmatrix}1\\\vct r_i\end{pmatrix}
 \begin{pmatrix}1\\\vct r_i\end{pmatrix}^{\!\top}.
 \label{eq:information}
\end{equation}
The entry $i_{\kappa\kappa}$ measures information about $\kappa$ when $\vtheta$ is known. When we estimate $\vtheta$, some of the variation in the score for $\kappa$ can also be explained by the nuisance scores. We subtract that linear prediction and use the variance of the remaining score, divided by $n$. Partitioning $\mat I$ according to $(\kappa,\vtheta^\top)^\top$, this gives the normalized efficient information
\begin{equation}
 j_{\mathrm{eff}} = i_{\kappa\kappa}-\vct i_{\theta\kappa}^\top
                  \mat I_{\theta\theta}^{-1}\vct i_{\theta\kappa}.
 \label{eq:effective}
\end{equation}
The subtracted term is the information lost through estimating $\vtheta$; the resulting matrix expression is the Schur complement. The same quantity has a weighted least-squares form:
\begin{equation}
 j_{\mathrm{eff}} = \min_{\vct q\in\R^p}\sum_i\frac{\bar e_i}{h_i}\{1-\vct r_i^\top\vct q\}^2.
 \label{eq:projection}
\end{equation}
Expanding the squared residuals in Eq.~\eqref{eq:projection} gives
\begin{equation*}
 i_{\kappa\kappa}-2\vct q^\top\vct i_{\theta\kappa}
 +\vct q^\top\mat I_{\theta\theta}\vct q.
\end{equation*}
Its minimizing coefficient vector is $\mat I_{\theta\theta}^{-1}\vct i_{\theta\kappa}$. Substituting this vector gives Eq.~\eqref{eq:effective}.

The constant vector represents an equal increase in the force of mortality at every age. The derivative columns represent the changes produced by the nuisance parameters. Eq.~\eqref{eq:projection} measures the weighted difference left after approximating the first with the second. If the nuisance information matrix is nonsingular, $j_{\mathrm{eff}}>0$ exactly when the derivative columns cannot reproduce the constant vector. This local requirement is separate from the global identification condition in (\ref{ass:identification}).

\section{\texorpdfstring{Proof of Theorem~\ref{thm:general}}{Proof of the main theorem}}
\label{app:proof}

We establish score normality, consistency, and the local quadratic expansion, then maximize the quadratic over $u\geq0$. This retains only the positive part of the efficient score and produces the point mass at zero. The remaining subsections derive local power and connect the argument to the profile quantities $U$ and $J$ of Section~\ref{sec:null}.

\subsection{Score, consistency, and local expansion}
\label{app:lan}

We collect the parameters in $\vzeta = (\kappa,\vtheta^\top)^\top$ and denote their true values under $H_0$ by $\vzeta_0$. We define the fixed vector
\begin{equation*}
 \vct b_i = \begin{pmatrix}1\\\vct r_i\end{pmatrix}.
\end{equation*}
The normalized random score vector is
\begin{equation*}
 \vct S_n = \sum_i\frac{D_{i,n}-n\bar e_i h_i}{\sqrt n\,h_i}\vct b_i.
\end{equation*}
Each count has a mean tending to infinity; centring and standardizing gives a normal limit. Independence across ages and the multivariate convergence argument \citep[Chapter~2]{van2000asymptotic} then give
\begin{equation}
 \vct S_n\dto\Normal_{p+1}(\vct 0,\mat I).
 \label{eq:normal}
\end{equation}

For consistency, we first use $D_{i,n}/n\pto\bar e_i h_i$, the fixed number of ages, and the bounds in (\ref{ass:identification}). Together with continuity and compactness, these give uniform convergence of $n^{-1}\{\ell_n(\vzeta)-\ell_n(\vzeta_0)\}$ to
\begin{equation*}
 \ell_\infty(\vzeta) = \sum_i\bar e_i\left[
 h_i\log\left\{\frac{ \bar{\mu}_i(\vzeta)}{h_i}\right\}
 -\{ \bar{\mu}_i(\vzeta)-h_i\}\right].
\end{equation*}
Each summand is nonpositive and equals zero only when $\bar\mu_i(\vzeta)=h_i$. Identification makes $\vzeta_0$ the unique maximizer. Uniform convergence and (\ref{ass:maximization}) give consistency of both estimators by the argmax consistency argument \citep[Chapter~5]{van2000asymptotic}.

At the true parameter, $\mat I$ is the probability limit of the negative Hessian divided by $n$. Terms involving second derivatives multiply centered count residuals and vanish after normalization. The derivative continuity in (\ref{ass:smoothness-information}) gives the corresponding local convergence. Positive definiteness means that the expected log-likelihood decreases quadratically in every parameter direction away from the true value. Its curvature grows at rate $n$, whereas the score is $\Op(\sqrt n)$. Together with consistency, these orders imply that the estimation errors are $\Op(n^{-1/2})$; the boundary restriction changes the admissible directions, not this rate \citep{self1987asymptotic}.

We express parameter displacements on this scale by defining
\begin{equation*}
 u=\sqrt n\,\kappa, \qquad
 \vct t=\sqrt n\,(\vtheta-\vtheta_0).
\end{equation*}
We hold $u\geq0$ and $\vct t\in\R^p$ fixed as $n$ increases, so the corresponding parameters approach the null at rate $n^{-1/2}$. Writing $S_{\kappa,n}$ and $\vct S_{\theta,n}$ for the corresponding blocks of $\vct S_n$, Taylor expansion gives, uniformly for bounded displacements,
\begin{equation}
\ell_n\!\left(\frac{u}{\sqrt n},
\vtheta_0+\frac{\vct t}{\sqrt n}\right)
-\ell_n(0,\vtheta_0)
=uS_{\kappa,n}
+\vct t^\top\vct S_{\theta,n} 
-\frac12\left(
u^2i_{\kappa\kappa}
+2u\,\vct i_{\theta\kappa}^\top\vct t
+\vct t^\top\mat I_{\theta\theta}\vct t
\right) 
+\op(1).
\label{eq:lan}
\end{equation}
Bounds on third derivatives control the remainder. The $\Op(n^{-1/2})$ estimation errors allow us to evaluate this expansion at the estimators, whose scaled displacements remain bounded in probability.

\subsection{Maximization under the boundary constraint}
\label{app:boundary}

Let $Q_n(u,\vct t)$ denote the quadratic expression on the right-hand side of Eq.~\eqref{eq:lan}, excluding the remainder $\op(1)$. For fixed $u$, maximizing this quadratic over $\vct t$ gives the random vector
\begin{equation*}
 \vct T_*(u)  =  \mat I_{\theta\theta}^{-1}
             (\vct S_{\theta,n}-\vct i_{\theta\kappa}u).
\end{equation*}
The nuisance maximizer of $Q_n$ at $u=0$ is $\vct T_*(0)$. We define the gain over this null maximum by
\begin{equation*}
 q_n(u)=Q_n\{u,\vct T_*(u)\}-Q_n\{0,\vct T_*(0)\}.
\end{equation*}
The corresponding log-likelihood difference is
\begin{equation}
 \ell_n\left(\frac{u}{\sqrt n},\vtheta_0+\frac{\vct T_*(u)}{\sqrt n}\right)
 -\ell_n\left(0,\vtheta_0+\frac{\vct T_*(0)}{\sqrt n}\right) =q_n(u)+\op(1).
 \label{eq:local-profile-gain}
\end{equation}
Substituting the two maximizers into $Q_n$ cancels the common nuisance-only term and gives
\begin{equation*}
 q_n(u)=uW_n-\tfrac12j_{\mathrm{eff}}u^2,
\end{equation*}
where the normalized efficient score is
\begin{equation*}
 W_n = S_{\kappa,n}-\vct i_{\theta\kappa}^\top
                    \mat I_{\theta\theta}^{-1}\vct S_{\theta,n}.
\end{equation*}
This subtracts the linear projection of the score for $\kappa$ on the nuisance scores. From Eq.~\eqref{eq:normal},
\begin{equation*}
 W_n\dto\Normal(0,j_{\mathrm{eff}}).
\end{equation*}
Over $u\in\R$ the quadratic would be maximized at $W_n/j_{\mathrm{eff}}$, of either sign, whereas over $u\geq0$ the maximizer is
\begin{equation*}
 U_* = \frac{\pos{W_n}}{j_{\mathrm{eff}}},
\end{equation*}
which is zero whenever $W_n\leq0$. Twice the maximized likelihood difference is
\begin{equation*}
 T = \frac{\pos{W_n}^{\,2}}{j_{\mathrm{eff}}}+\op(1),
\end{equation*}
so that, for a standard normal $Z_0$,
\begin{equation*}
 T\dto\pos{Z_0}^{\,2}.
\end{equation*}
The right-hand side is zero when $Z_0\leq0$, an event of probability one half, and conditional on $Z_0>0$ its square is chi-squared with one degree of freedom. This establishes Theorem~\ref{thm:general} and its rejection rule.

\subsection{Local power}
\label{app:power}

The local log-likelihood expansion Eq.~\eqref{eq:lan} gives the power of the test, as \citet{bohnstedt2019detecting} show for the frailty variance of the gamma-Gompertz model. Under local alternatives $\kappa_n=u/\sqrt n$ with $u>0$ fixed, Le Cam's third lemma gives $W_n\dto\Normal(u\,j_{\mathrm{eff}},j_{\mathrm{eff}})$ \citep{van2000asymptotic}, so that $T\dto\pos{Z_0+u\sqrt{j_{\mathrm{eff}}}}^{\,2}$. Because $\chi^2_{1,1-2\alpha}=z_{1-\alpha}^2$, the rule in Eq.~\eqref{eq:rule} rejects with probability tending to $\mathbb P(Z_0+u\sqrt{j_{\mathrm{eff}}}>z_{1-\alpha})=\Phi(u\sqrt{j_{\mathrm{eff}}}-z_{1-\alpha})$.

For a target power $\pi\in(\alpha,1)$, inverting the preceding limiting power gives
\begin{equation*}
 u\sqrt{j_{\mathrm{eff}}}=z_{1-\alpha}+z_\pi.
\end{equation*}
Substituting $\kappa=u/\sqrt n$ gives the local detection approximation in Section~\ref{sec:power}. We evaluate the information at the null, so this inversion does not give an exact finite-sample threshold.

\subsection{Connection to the profile score}
\label{app:profile}

To connect this with the profile quantities of Section~\ref{sec:null}, note first that $\vtheta$ sits at its unconstrained maximum inside $\ell_p$, so differentiating that maximization with respect to $\kappa$ contributes nothing and the profile score $U$ equals the ordinary score for $\kappa$ evaluated at the restricted estimator $\widehat{\vtheta}_{0,n}$. Expanding its nuisance score equation gives
\begin{equation*}
 \sqrt n(\widehat{\vtheta}_{0,n}-\vtheta_0)
  = \mat I_{\theta\theta}^{-1}\vct S_{\theta,n}+\op(1).
\end{equation*}
Let $\mat V_n$ denote the observed information matrix, the negative Hessian of $\ell_n$ at $(0,\widehat{\vtheta}_{0,n})$, with scalar entry $v_{\kappa\kappa,n}$, vector block $\vct v_{\theta\kappa,n}$ and matrix block $\mat V_{\theta\theta,n}$. Differentiating the nuisance score equation along the profile gives
\begin{equation*}
 J = v_{\kappa\kappa,n}-\vct v_{\theta\kappa,n}^\top\mat V_{\theta\theta,n}^{-1}\vct v_{\theta\kappa,n}.
\end{equation*}
Combining $\mat V_n/n\pto\mat I$ with the score expansion,
\begin{equation*}
 n^{-1/2}U = W_n+\op(1), \qquad
 J/n \pto j_{\mathrm{eff}}.
\end{equation*}
These limits justify the standardization Eq.~\eqref{eq:zlimit} and the quadratic approximation in Eq.~\eqref{eq:gmquadratic}, and in particular they are what permits the fitted residuals at $\widehat{\vtheta}_{0,n}$, which are not independent across ages, to be replaced by the independent residuals at $\vtheta_0$, to which a central limit theorem applies.

\section{Local information conditions}
\label{app:info}

Theorem~\ref{thm:general} requires that no combination of nuisance-parameter changes can mimic a constant addition to mortality. We verify that condition for the two models of Section~\ref{sec:test} and then bound the information available on a narrow age window.

\subsection{Gompertz--Makeham} \label{app:gm}

We show that the condition holds as soon as three distinct ages are observed with positive exposure. Mimicking the constant would require an exponential, a constant, and a linear function of age to be linearly dependent at three points, which Rolle's theorem excludes.

We use the nuisance parameter vector $\vtheta = (\log\lambda,\beta)^\top$. Under $H_0$, $h_i = \lambda e^{\beta x_i}$ and
\begin{equation*}
 \vct r_i = \begin{pmatrix}h_i\\x_i h_i\end{pmatrix}.
\end{equation*}
The blocks of the normalized Fisher information matrix are
\begin{equation*}
 i_{\kappa\kappa} = \sum_i\bar e_i/h_i,\qquad
 \vct i_{\theta\kappa} =
 \begin{pmatrix}\sum_i\bar e_i\\\sum_i\bar e_i x_i\end{pmatrix},\qquad
 \mat I_{\theta\theta} =
 \begin{pmatrix}
 \sum_i\bar e_i h_i & \sum_i\bar e_i x_i h_i\\
 \sum_i\bar e_i x_i h_i & \sum_i\bar e_i x_i^2 h_i
 \end{pmatrix}.
\end{equation*}
A positive definite Fisher information matrix for three parameters requires at least three distinct ages with positive exposures. To see why three are enough when $\lambda,\beta>0$, consider the derivative matrix
\begin{equation*}
 \mat B=
 \begin{pmatrix}
 1&h_1&x_1h_1\\
 \vdots&\vdots&\vdots\\
 1&h_m&x_mh_m
 \end{pmatrix}.
\end{equation*}
Its columns correspond to changes in $\kappa$, $\log\lambda$, and $\beta$. If they were linearly dependent, constants $\rho_1,\rho_2,\rho_3$, not all zero, would satisfy
\begin{equation*}
 \lambda\rho_1+\rho_2h_i+\rho_3x_ih_i=0
\end{equation*}
at every fitted age. The factor $\lambda>0$ in the first coefficient only rescales that coefficient. Dividing by $h_i=\lambda e^{\beta x_i}$ gives
\begin{equation*}
 f(x_i)=\rho_1e^{-\beta x_i}+\rho_2+\rho_3x_i=0.
\end{equation*}
Linear dependence would require the function $f$ to have zeros at three distinct ages. By Rolle's theorem, three zeros of $f$ require at least two zeros of $f'$ and hence at least one zero of $f''$. But
\begin{equation*}
 f''(x)=\rho_1\beta^2e^{-\beta x}
\end{equation*}
never equals zero if $\rho_1\neq0$. If $\rho_1=0$, $f$ is linear, and three zeros force $\rho_2=\rho_3=0$. Both cases contradict the proposed dependence. The columns have full rank. Positive exposure weights give a positive definite Fisher information matrix and $j_{\mathrm{eff}}>0$. With the remaining assumptions in Appendix~\ref{app:regularity}, Theorem~\ref{thm:general} establishes Proposition~\ref{prop:gm}.

\subsection{Information on a narrow age window}
\label{app:shortwindow}

The preceding proof shows that the Makeham term is locally distinguishable from level and slope changes. It does not show how well we can distinguish them. We now show why that distinction becomes weak on a narrow age window, supplying the bound behind Section~\ref{sec:power}. Write $t=x-x_0$ and $h_0=\lambda e^{\beta x_0}$. The function
\begin{equation*}
 \frac{1+\beta x_0}{h_0}h(x)-\frac{\beta}{h_0}xh(x)=(1-\beta t)e^{\beta t}
\end{equation*}
is a linear combination of $h$ and $xh$. Taylor expansion gives
\begin{equation*}
 (1-\beta t)e^{\beta t}=1-\tfrac12\beta^2t^2+O(t^3).
\end{equation*}
This supplies a candidate in Eq.~\eqref{eq:projection} whose residual is of order $t^2$. On a window with $|t|\leq w$, for fixed $\beta$ and small $w$, it gives
\begin{equation*}
 j_{\mathrm{eff}}\leq Cw^4\sum_i\bar e_i/h_i
\end{equation*}
for some constant $C$ independent of $w$. A narrow window makes the constant difficult to distinguish from level and slope adjustments when the total information weight remains bounded. This local bound does not specify the information loss for every change of age range or exposure pattern.

\subsection{Gamma-Gompertz--Makeham}
\label{app:ggm}

We now supply the derivatives of the gamma-Gompertz hazard needed to form the Fisher information matrix when frailty is present, and then show that four distinct ages with positive exposures give that matrix full rank whenever $\lambda\gamma\neq\beta$. As in Appendix~\ref{app:gm}, the argument turns on the linear independence of a small set of functions of age, and what it establishes is the local information condition rather than global identification.

We use $\vtheta = (\log\lambda,\beta,\gamma)^\top$ and write $g$ for the hazard in Eq.~\eqref{eq:gg}. Define the auxiliary functions
\begin{equation*}
 q(x)=\frac\lambda\beta(e^{\beta x}-1), \qquad
 s(x)=1+\gamma q(x), \qquad
 q_\beta(x)=\frac{\lambda}{\beta^2}\{\beta x e^{\beta x}-(e^{\beta x}-1)\}.
\end{equation*}
Keeping the age argument explicit, differentiation gives
\begin{equation*}
 \frac{\partial g(x)}{\partial\log\lambda}=\frac{g(x)}{s(x)},\qquad
 \frac{\partial g(x)}{\partial\beta}=g(x)\left\{x-\frac{\gamma q_\beta(x)}{s(x)}\right\}, \qquad
 \frac{\partial g(x)}{\partial\gamma}=-\frac{g(x)q(x)}{s(x)}.
\end{equation*}
Substituting these into Eq.~\eqref{eq:information} determines the normalized Fisher information matrix. When $\gamma_0>0$, the frailty variance can be both increased and decreased by a sufficiently small amount, and the other nuisance parameters satisfy (\ref{ass:parameter-space}). We exclude $\lambda\gamma = \beta$, because it makes $g$ constant and violates regular identification of the added constant.

Four distinct ages with positive exposures then suffice for full rank. Write $A = 1-\gamma\lambda/\beta$ and $B = \gamma\lambda/\beta$, so that $s(x) = A+Be^{\beta x}$, with $A\neq0$ precisely when $\lambda\gamma\neq\beta$. Multiplying row $i$ by the positive factor $s(x_i)^2/(\lambda e^{\beta x_i})$ leaves the rank unchanged and turns the four derivative columns for $(\kappa,\log\lambda,\beta,\gamma)$ into
\begin{align*}
 \kappa&\longmapsto \frac{A^2}{\lambda}e^{-\beta x}+\frac{2AB}{\lambda}+\frac{B^2}{\lambda}e^{\beta x},\\
 \log\lambda&\longmapsto 1,\\
 \beta&\longmapsto Ax+\frac{B}{\beta}\left(e^{\beta x}-1\right),\\
 \gamma&\longmapsto \frac{\lambda}{\beta}\left(1-e^{\beta x}\right),
\end{align*}
where the terms in $xe^{\beta x}$ cancel in the third column. Using rows in the order $e^{-\beta x},1,x,e^{\beta x}$ and columns in the order $\kappa,\log\lambda,\beta,\gamma$, their coefficients form the matrix
\begin{equation*}
 \mat C=
 \begin{pmatrix}
 A^2/\lambda&0&0&0\\
 2AB/\lambda&1&-B/\beta&\lambda/\beta\\
 0&0&A&0\\
 B^2/\lambda&0&B/\beta&-\lambda/\beta
 \end{pmatrix}.
\end{equation*}
Expansion along its first row and then the row containing $A$ gives
\begin{equation*}
 \det\mat C=\frac{A^2}{\lambda}\,A\left(-\frac{\lambda}{\beta}\right)=-\frac{A^3}{\beta}.
\end{equation*}
The transformation between these basis functions and the derivative columns is invertible exactly when $\lambda\gamma\neq\beta$.

It remains to check that the four basis functions are linearly independent when evaluated at four distinct ages $x_1<x_2<x_3<x_4$. If they were dependent, a linear combination with coefficients not all zero,
\begin{equation*}
 f(x)=\rho_1e^{-\beta x}+\rho_2+\rho_3x+\rho_4e^{\beta x},
\end{equation*}
would satisfy $f(x_i)=0$ for $i=1,2,3,4$. Rolle's theorem would then give at least three zeros of $f'$ between these ages and at least two zeros of $f''$. However,
\begin{equation*}
 f''(x) = \beta^2\left(\rho_1e^{-\beta x}+\rho_4e^{\beta x}\right)
\end{equation*}
has at most one unless $\rho_1 = \rho_4 = 0$, in which case $f$ is linear and its four zeros force $\rho_2 = \rho_3 = 0$. The derivative matrix has full rank, positive exposure weights give a positive definite normalized Fisher information matrix, and $j_{\mathrm{eff}}>0$. This covers $\gamma_0 = 0$ as well as $\gamma_0>0$. As in Appendix~\ref{app:gm} the result is the local information condition, with global identification a separate requirement in (\ref{ass:identification}). When $\gamma_0>0$ and the remaining assumptions hold, Theorem~\ref{thm:general} establishes Proposition~\ref{prop:ggm}.

\section{Distribution under \texorpdfstring{$H_0$}{H0} when the true frailty variance is zero}
\label{app:zero-frailty}

When the true frailty variance is zero, two parameters sit on boundaries at once, changing the set of admissible local parameter displacements. Instead of maximizing a quadratic over a half-line, we maximize it over a quadrant under the alternative and over one edge of that quadrant under the null.

\subsection{Limiting statistic}
\label{app:zero-limit}

At a true parameter with $\kappa = \gamma = 0$, we collect the constrained parameters in $\vpsi = (\kappa,\gamma)^\top$ and the nuisance parameters in $\veta = (\log\lambda,\beta)^\top$. The vector $\veta$ can be varied in every direction by a sufficiently small amount. We assume the analogues of (\ref{ass:parameter-space})--(\ref{ass:maximization}) with both nonnegativity constraints. Normalizing the Fisher information matrix by $n$, as in Eq.~\eqref{eq:information}, and adjusting for estimation of $\veta$ gives
\begin{equation}
 \mat K = \mat I_{\psi\psi}
       -\mat I_{\psi\eta}\mat I_{\eta\eta}^{-1}\mat I_{\eta\psi},
 \label{eq:kmatrix}
\end{equation}
which we assume positive definite. At $\kappa=\gamma=0$, define the derivative vector in the order $(\kappa,\gamma,\log\lambda,\beta)^\top$ by
\begin{equation*}
 \vct d_i=(1,-h_iq_i,h_i,x_ih_i)^\top,
\end{equation*}
where $q_i=q(x_i)$ as in Appendix~\ref{app:ggm}. For this parameter order, Eq.~\eqref{eq:information} becomes
\begin{equation*}
 \mat I=\sum_i\frac{\bar e_i}{h_i}\vct d_i\vct d_i^\top.
\end{equation*}
We partition this matrix into the first two and last two coordinates, then substitute its blocks into Eq.~\eqref{eq:kmatrix} to obtain $\mat K$.

Applying the score limit Eq.~\eqref{eq:normal} to this four-parameter vector gives $\vct S\sim\Normal_4(\vct 0,\mat I)$. Write $\vct S=(\vct S_\psi^\top,\vct S_\eta^\top)^\top$, so $\vct S_\psi$ contains the scores for $\kappa$ and $\gamma$, and $\vct S_\eta$ contains the nuisance scores for $\log\lambda$ and $\beta$. The best linear prediction of $\vct S_\psi$ from $\vct S_\eta$ is $\mat I_{\psi\eta}\mat I_{\eta\eta}^{-1}\vct S_\eta$, using their covariance blocks. Subtracting this prediction gives
\begin{equation*}
 \vct W = \vct S_\psi-\mat I_{\psi\eta}\mat I_{\eta\eta}^{-1}\vct S_\eta,
\end{equation*}
so that
\begin{equation*}
 \vct W\sim\Normal_2(\vct 0,\mat K).
\end{equation*}
Let $\vct z=(z_1,z_2)^\top$ denote the fixed local displacements of $\vpsi$, so $\kappa=z_1/\sqrt n$ and $\gamma=z_2/\sqrt n$. After maximizing over the local displacements of $\veta$, we obtain the quadratic
\begin{equation*}
 q_{\vct W}(\vct z) = \vct z^\top\vct W-\tfrac12\vct z^\top\mat K\vct z.
\end{equation*}
Under the full model we maximize over $\vct z\in\R_+^2$, and under $H_0$ we set $z_1 = 0$ and maximize over $z_2\geq0$. Using the expansion in Eq.~\eqref{eq:lan} with both constraints,
\begin{equation*}
 T\dto T_0(\vct W),
\end{equation*}
where
\begin{equation*}
 T_0(\vct W)
  = 2\left\{\sup_{\vct z\in\R_+^2}q_{\vct W}(\vct z)
          -\sup_{t\geq0}q_{\vct W}\left((0,t)^\top\right)\right\}.
\end{equation*}
In the first supremum both local parameters are free to be nonnegative, while in the second the local value of $\kappa$ is fixed at zero and the local value of $\gamma$ is estimated. This is the geometry of a quadrant against one of its edges, as in the nuisance-boundary cases of \citet{self1987asymptotic}, and because the null is an edge rather than the origin, the difference of the two suprema is not in general distributed as a mixture of chi-squared variables.

\subsection{Explicit form and critical values}
\label{app:zero-explicit}

The constrained Gaussian quadratic follows from the boundary likelihood theory of \citet[Theorem~3]{self1987asymptotic}. We evaluate it here for the quadrant and its null edge by considering the origin, the two edges, and the interior. Writing $k_{ij}$ for the entries of $\mat K$, the candidate values for twice the full maximum are
\begin{equation*}
 A_1 = \pos{W_1}^{\,2}/k_{11}, \qquad
 A_2 = \pos{W_2}^{\,2}/k_{22},
\end{equation*}
corresponding to the two edges of the quadrant, together with
\begin{equation*}
 A_3 = \vct W^\top\mat K^{-1}\vct W
\end{equation*}
when $\mat K^{-1}\vct W\in\R_+^2$, corresponding to an admissible unconstrained maximizer of the strictly concave quadratic. The origin needs no separate candidate, because $A_1$ and $A_2$ are already nonnegative and equal zero exactly when the corresponding edge maximum is at the origin. Collecting the admissible candidates in $\mathcal A(\vct W)$,
\begin{equation}
 T_0(\vct W) = \max\mathcal A(\vct W)-A_2,
 \label{eq:zero-frailty-explicit}
\end{equation}
where subtracting $A_2$ removes the maximized contribution under $H_0$.

Eq.~\eqref{eq:zero-frailty-explicit} is what makes the rejection rule in Eq.~\eqref{eq:zero-frailty-rule} usable. Fit the Gompertz model with $\kappa$ and $\gamma$ both fixed at zero, form $\widehat{\mat K}$ from the resulting $\widehat\lambda,\widehat\beta$ using Eq.~\eqref{eq:information} and the partition in Eq.~\eqref{eq:kmatrix}, draw $B$ replicates $\vct W\sim\Normal_2(\vct 0,\widehat{\mat K})$, evaluate $T_0(\vct W)$ at each draw, and take the empirical $1-\alpha$ quantile as the critical value. Two approximations enter. The limit is asymptotic, and $\widehat{\mat K}$ replaces $\mat K$, which needs the Gompertz fit to be consistent under $\kappa = \gamma = 0$. The $B$ draws add Monte Carlo error to the quantile, which can be made small by taking $B$ large.

\subsection{\texorpdfstring{The role of $k_{12}$}{The role of k12}}
\label{app:zero-negcorr}

\paragraph*{Uncorrelated adjusted scores.}
If $\mat K$ is diagonal, the quadratic separates into contributions from $\kappa$ and $\gamma$. The maximized contribution from $\gamma$ is then the same under both hypotheses and cancels in the likelihood ratio, leaving
\begin{equation*}
 T_0(\vct W) = \pos{W_1}^{\,2}/k_{11},
\end{equation*}
so the limit reverts to $\tfrac12\delta_0+\tfrac12\chi_1^2$. Otherwise the fits of the two constrained parameters interact, the cancellation fails, and the distribution must be determined from Eq.~\eqref{eq:zero-frailty-explicit}. In particular, the critical value $2.7055$, which is the 95th percentile of $\tfrac12\delta_0+\tfrac12\chi_1^2$, cannot be assumed to be the 95th percentile of $F_{\mat K}$.

\paragraph*{Negatively correlated adjusted scores.}
When $k_{12}<0$, we can establish the direction of the error. We first derive the statistic for each score configuration, then calculate its mass at zero and compare its upper tail with the half-and-half law. Write $\rho=k_{12}/\sqrt{k_{11}k_{22}}$, and let
\begin{equation*}
 W_{1\cdot2} =W_1-\frac{k_{12}}{k_{22}}W_2, \qquad
 k_{11\cdot2} =k_{11}-\frac{k_{12}^2}{k_{22}},
\end{equation*}
with $W_{2\cdot1}$ and $k_{22\cdot1}$ defined by exchanging the indices. Then $W_{1\cdot2}\sim\Normal(0,k_{11\cdot2})$ is independent of $W_2$, and
\begin{equation*}
 \vct W^\top\mat K^{-1}\vct W =\frac{W_2^2}{k_{22}}+\frac{W_{1\cdot2}^2}{k_{11\cdot2}}, \qquad
 \vct W^\top\mat K^{-1}\vct W =\frac{W_1^2}{k_{11}}+\frac{W_{2\cdot1}^2}{k_{22\cdot1}}.
\end{equation*}
The components of $\vct z=\mat K^{-1}\vct W$ are $z_1=W_{1\cdot2}/k_{11\cdot2}$ and $z_2=W_{2\cdot1}/k_{22\cdot1}$, and they satisfy $k_{12}z_1+k_{22}z_2=W_2$.

\paragraph*{Case 1: $W_2>0$.}
If $W_{1\cdot2}>0$, the unconstrained maximizer has $z_1>0$. Because $k_{12}<0$, its other coordinate also satisfies
\begin{equation*}
 z_2=\frac{W_2-k_{12}z_1}{k_{22}}>0.
\end{equation*}
Both coordinates are admissible, so the full maximum is $A_3$. Subtracting the null maximum gives
\begin{equation*}
 T_0=A_3-A_2=\frac{W_{1\cdot2}^2}{k_{11\cdot2}}.
\end{equation*}
If instead $W_{1\cdot2}\leq0$, then $W_1\leq(k_{12}/k_{22})W_2<0$. The first edge gives $A_1=0$, and the interior cannot improve on the null edge. Hence $T_0=0$.

\paragraph*{Case 2: $W_2\leq0$.}
Here the null maximum is $A_2=0$. If $W_1\leq0$ as well, both linear terms in the quadratic are nonpositive on the quadrant. The full maximum is at the origin, and $T_0=0$.

If $W_1>0$, the first edge gives $A_1=W_1^2/k_{11}$. The interior maximum is admissible exactly when $W_{2\cdot1}\geq0$. To check the other coordinate in that event, observe that
\begin{equation*}
 k_{22}W_1-k_{12}W_2\geq\frac{W_1\det\mat K}{k_{11}}>0.
\end{equation*}
Then $z_1>0$ too. The second quadratic decomposition then adds $W_{2\cdot1}^2/k_{22\cdot1}$ to $A_1$; otherwise the first edge remains the maximum. Combining the cases gives
\begin{equation*}
 T_0=
 \begin{cases}
  \pos{W_{1\cdot2}}^{2}/k_{11\cdot2}, & W_2>0,\\[2pt]
  W_1^2/k_{11}+\pos{W_{2\cdot1}}^{2}/k_{22\cdot1}, & W_2\leq0,\ W_1>0,\\[2pt]
  0, & W_2\leq0,\ W_1\leq0.
 \end{cases}
\end{equation*}

\paragraph*{Probability mass at zero.}
Sheppard's formula for the orthant probabilities of a bivariate normal distribution gives $\mathbb P(W_1\leq0,W_2\leq0)=\tfrac14+\arcsin(\rho)/(2\pi)$ for any correlation $\rho$ \citep{sheppard1899application}. Together with $\mathbb P(W_2>0,W_{1\cdot2}\leq0)=\tfrac14$, which holds because $W_2$ and $W_{1\cdot2}$ are independent and centered, the piecewise form gives
\begin{equation*}
 \mathbb P(T_0=0)=\frac12+\frac{\arcsin\rho}{2\pi}<\frac12.
\end{equation*}

\paragraph*{Upper-tail comparison.}
For the upper tail, fix $c>0$, let $V\sim\chi^2_1$ and $a=\sqrt{c\,k_{11}}$, so that $\tfrac12\mathbb P(V>c)=\mathbb P(W_1>a)$. By the independence of $W_{1\cdot2}$ and $W_2$, the first case contributes $\mathbb P(T_0>c,\,W_2>0)=\tfrac14\mathbb P(V>c)=\tfrac12\mathbb P(W_1>a)$. In the second case $T_0\geq W_1^2/k_{11}$, so it contributes at least $\mathbb P(W_2\leq0,\,W_1>a)$. Therefore
\begin{equation*}
\mathbb P(T_0>c)-\tfrac12\mathbb P(V>c) \geq \mathbb P(W_1>a) 
\left\{ \mathbb P(W_2\leq0\mid W_1>a)-\tfrac12
\right\}>0.
\end{equation*}
because, given $W_1=w>0$, $W_2$ is normal with mean $(k_{12}/k_{11})w<0$. Taking $c=\chi^2_{1,1-2\alpha}$ shows that the rule in Eq.~\eqref{eq:rule} has asymptotic rejection probability above $\alpha$ at every level. When $k_{12}=0$ the bound holds with equality, in agreement with the case of uncorrelated adjusted scores.

\subsection{Frailty variances near zero}
\label{app:zero-near}

For frailty variances close to but not equal to zero, set $\kappa = 0$ and $\gamma_n = \tau/\sqrt n$ with fixed $\tau\geq0$. Expanding the log-likelihood about $\kappa = \gamma = 0$, the limiting adjusted score is
\begin{equation*}
 \vct W_\tau\sim\Normal_2\left(\mat K(0,\tau)^\top,\mat K\right),
\end{equation*}
and the statistic satisfies
\begin{equation*}
 T\dto T_0(\vct W_\tau).
\end{equation*}
The limiting distribution can depend on $\tau$, which describes how the true frailty variance approaches zero as $n\to\infty$ under Eq.~\eqref{eq:exposure}. The result at exactly $\gamma = 0$ and the result at a fixed $\gamma>0$ do not by themselves bound the rejection probability across all such sequences. Controlling the type~I error when $\gamma$ is unknown requires an argument that applies uniformly across these values, and selecting a reference distribution according to whether $\widehat{\gamma} = 0$ does not supply one. The results remain valid when frailty is fixed at zero by specification, or when its true variance is fixed and positive.



\printbibliography


\end{document}